\documentclass[aps,twocolumn,superscriptaddress,showpacs,amssymb,prl,floatfix,longbibliography]{revtex4-1}

\usepackage{graphicx,color}
\usepackage{dcolumn}% Align table columns on decimal point
\usepackage{amsmath}
\usepackage{amssymb}
\usepackage{epstopdf}

\usepackage[]{hyperref}
\hypersetup{colorlinks=true,linkcolor=blue,citecolor=blue,urlcolor=blue,pdfpagemode=UseNone}

\begin{document}

\thispagestyle{myheadings}

%\title{Electric field gradient measurements of the local nematic susceptibility in BaFe$_2$As$_2$ under uniaxial stress}
\title{Local nematic susceptibility in stressed BaFe$_2$As$_2$ from NMR electric field gradient measurements}

\author{T. Kissikov}

\affiliation{Department of Physics, University of California, Davis, California
95616, USA}

\author{R. Sarkar}

\affiliation{Institute for Solid State Physics, TU Dresden, D-01069 Dresden, Germany}

\author{M. Lawson}

\author{B. T. Bush}

\affiliation{Department of Physics, University of California, Davis, California
95616, USA}

\author{E. I. Timmons}

\author{M. A. Tanatar}

\author{R. Prozorov}

\author{S. L. Bud'ko}

\author{P. C. Canfield}

\affiliation{Ames Laboratory U.S. DOE and Department of Physics and Astronomy,Iowa
State University, Ames, Iowa 50011, USA}

\author{R. M. Fernandes}

\affiliation{School of Physics and Astronomy, University of Minnesota, Minneapolis,
Minnesota 55455, USA}

\author{W. F. Goh}

\author{W. E. Pickett}

\author{N. J. Curro}

\affiliation{Department of Physics, University of California, Davis, California
95616, USA}
\date{\today}
%%%%%%%%%%%%%%%%%%%%%%%%%%%%%%%%%%%%%%%%%%%%%%%%%%%%%%%%%%%%%%%%%%%%%
\begin{abstract}

The electric field gradient (EFG) tensor at the $^{75}$As site couples to the orbital occupations of the As p-orbitals and is a sensitive probe of local %ferro-orbital ordering
nematicity in BaFe$_2$As$_2$.
%Because this tensor has the point group symmetry of the lattice, the EFG asymmetry parameter is a direct measure of the local nematicity.
We use nuclear magnetic resonance to  measure the nuclear quadrupolar splittings and find that the EFG asymmetry responds linearly to the presence of a strain field in the paramagnetic phase. We extract the nematic susceptibility from the slope of this linear response as a function of temperature and find that it diverges near the structural transition in agreement with other measures of the bulk nematic susceptibility. Our work establishes an alternative method to extract the nematic susceptibility which, in contrast to transport methods, can be extended inside the superconducting state.

\end{abstract}

\pacs{76.60.-k, 75.30.Mb,  75.25.Dk, 76.60.Es}

%\pacs{75.40.Gb, 76.60.-k, 75.50.Bb, 75.50.Lk, 75.60.-d}

%%%%%%%%%%%%%%%%%%%%%%%%%%%%%%%%%%%%%%%%%%%%%%%%%%%%%%%%%%%%%%%%%%%%%
%%%%%%%%%%%%%%%%%%%%%%%%%%%%%%%%%%%%%%%%%%%%%%%%%%%%%%%%%%%%%%%%%%%%%

\maketitle

The iron-based superconductors exhibit a complex interplay between orbital, electronic and lattice degrees of freedom.  In BaFe$_2$As$_2$ a nematic instability triggers a spontaneous orthorhombic distortion, ferro-orbital order, and spin-fluctuations anisotropy below $T_s = 135$ K in the absence of strain \cite{doping122review,FernandesNematicPnictides}.  This nematic phase breaks the $C_4$ tetragonal symmetry of the lattice, and is preceded by critical nematic fluctuations and divergent nematic susceptibility in the disordered phase \cite{FisherScienceNematic2012,NMRnematicStrainBa122PRB2016}.  In the nematic phase, the Fe $d_{xz}$ and $d_{yz}$ orbitals become non-degenerate, with an energy splitting on the order of 40 meV, and different occupation levels \cite{NaFeAsARPES2012}.  This phase is closely related to the stripe antiferromagnetic ordering of the Fe spins, which order either concomitantly with the nematic phase transition, or at a temperature $T_N$ only a few Kelvins below, as it is the case in BaFe$_2$As$_2$. As a result, many low energy experimental probes actually sense a complex interplay of the orbital, lattice, and magnetic degrees of freedom simultaneously, precluding quantitative analyses.

\begin{figure}[!t]
	\centering
	\includegraphics[width=\linewidth]{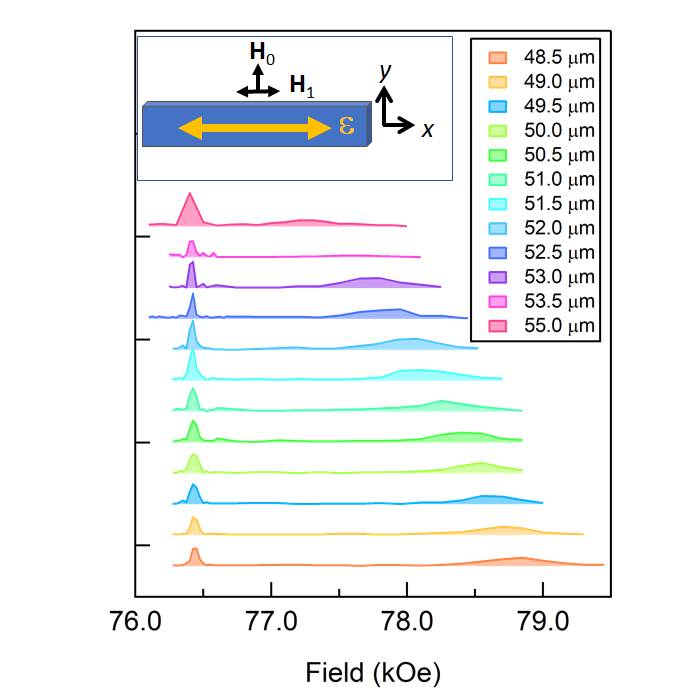}
\caption{\label{fig:waterfall} (color online) Field-swept spectra of BaFe$_2$As$_2$ at constant frequency $f=55.924$ MHz at 138 K for several different displacements of the piezoelectric device, showing the central and upper satellite transitions.  Zero strain corresponds to 51.58 $\mu$m. Inset: Orientation of the crystal with respect to the external field, $\mathbf{H}_0$, the strain axis, and the rf field $\mathbf{H}_1$. In this paper, $x$ and $y$ are parallel to the Fe-Fe directions.}
\end{figure}

Several techniques have been developed to probe the nematic degrees of freedom.
Anisotropic resistivity \cite{TanatarDetwin,IronArsenideDetwinnedFisherScience2010}, elastoresistance \cite{FisherScienceNematic2012}, electronic Raman scattering \cite{Ba122RamanPRL2013}, elastic constants \cite{NematicElasticPRL2010,IshidaFeSePRL2015,MeingastBaFe2As2strain2016,MeingastNematicSusceptibility2016}, thermopower \cite{ThermopowerNematicityPRL}, polarized light image color analysis \cite{TanatarTensileStressPRB,TanatarPRL2016} and optical conductivity \cite{Mirri2015} probe bulk anisotropies.
NMR and neutron scattering, on the other hand, and have been chiefly used to investigate the effect of nematicity on the spin fluctuations \cite{PhysRevB.89.214511,Kissikov2017,DaiRMP2015,StrainBa122neutronsPRB2015,StrainedPnictidesNS2014science}.  The nuclear quadrupolar interaction, however, can probe the microscopic orbital occupations directly \cite{IyeJPSLorbitalnematicity2015}.  The $^{75}$As ($I=3/2$) quadrupolar moment couples to the local electric field gradient (EFG), which is dominated by the on-site occupations of the As 4p electrons.  These orbitals are hybridized with the Fe 3d orbitals, and thus the EFG is a sensitive probe of the d-orbital occupations.  Indeed, the EFG tensor exhibits a dramatic lowering from axial symmetry at the nematic phase transition in the absence of applied strain %, as shown in Fig. \ref{fig:nuxx}
\cite{takigawa2008}.  In this Rapid Communication we present new data on the EFG under uniaxial strain applied in a controlled manner via a piezo device. We find that the EFG asymmetry parameter is linearly proportional to the in-plane strain applied to the crystal, and is thus a direct measure of the nematic susceptibility.   This approach enables one to probe the \emph{local}, rather than global, nematic susceptibility. Moreover, it in principle makes it possible to probe the nematic properties of the superconducting state, which is not accessible by elasto-resistance measurements.

%%%

\begin{figure}
	\centering
	\includegraphics[width=\linewidth]{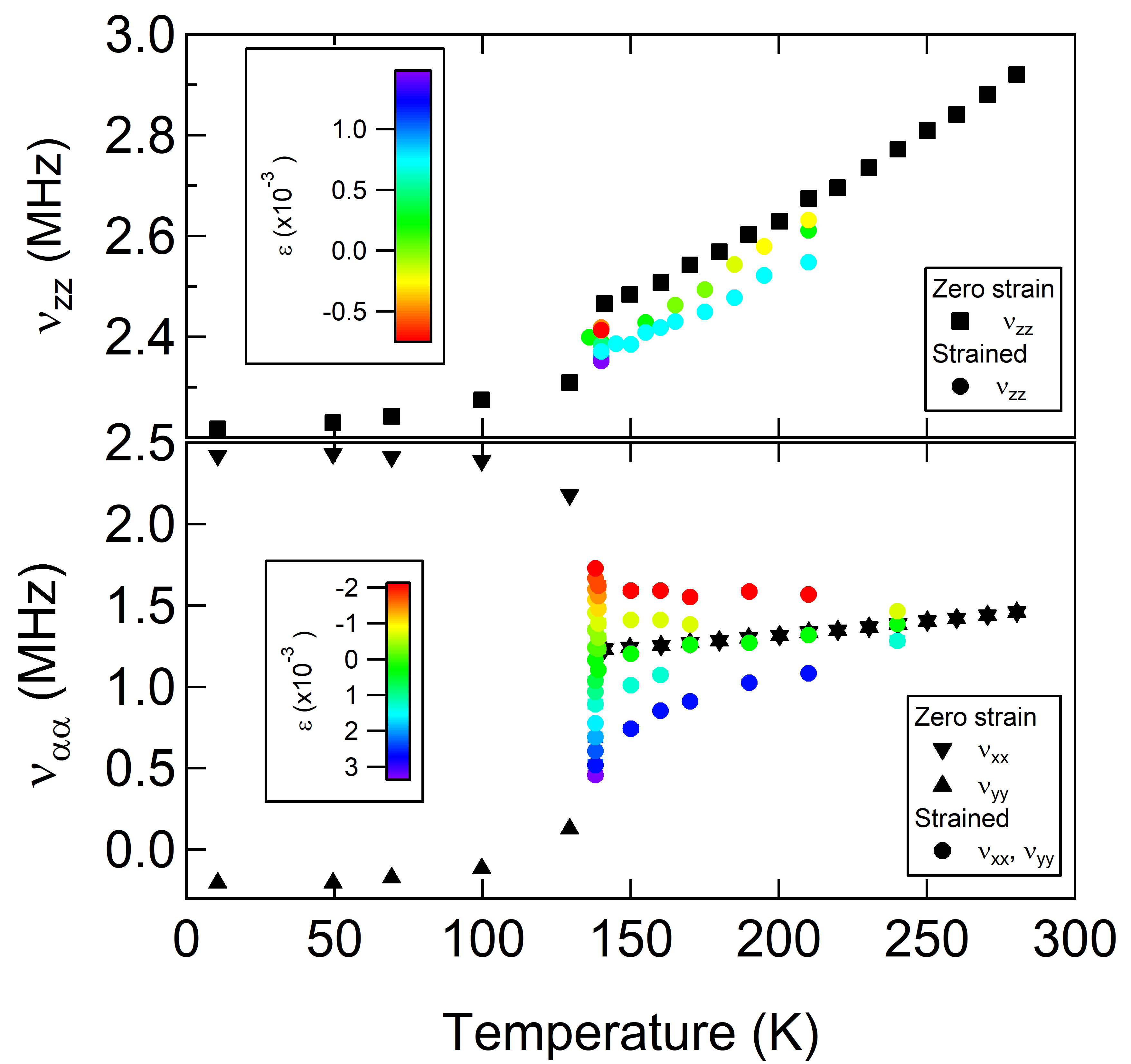}
\caption{\label{fig:nuxx} (color online) The As electric field gradient components ($\nu_{xx}$, $\nu_{yy}$, $\nu_{zz}$) versus temperature for BaFe$_2$As$_2$ both in zero strain (reproduced from \cite{takigawa2008}) and under uniaxial strain. %\textbf{\emph{RMF: I find the notation in this figure confusing, because $x$ and $y$ as introduced in the text are not parallel to $a$ and $b$. I think $\nu_{xx}$ should be replaced by $\nu_{aa}$ and $nu_{yy}$ by $nu_{bb}$.}}
}
\end{figure}

\begin{figure}
	\centering
	\includegraphics[width=\linewidth]{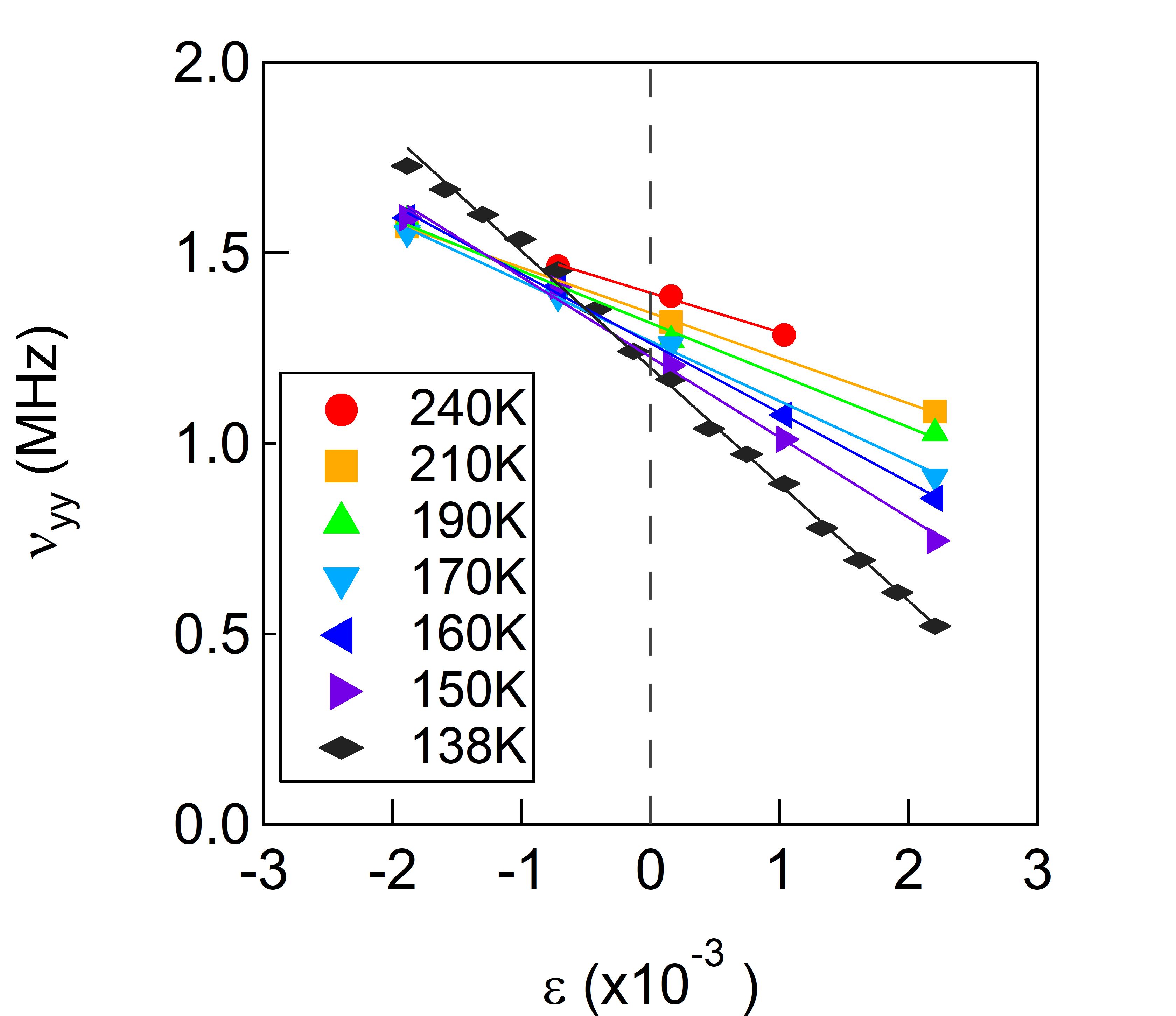}
	\caption{\label{fig:EFGlinear} (color online) The quadrupolar splitting $\nu_{yy}$ as a function of strain at several fixed temperatures.  The solid lines are linear fits to the data. }
\end{figure}

A single crystal of BaFe$_2$As$_2$ was synthesized via a self-flux method and cut to dimensions of approximately 1.5 mm$\times$0.5 mm with the long axis parallel to the (110)$_{T}$ direction in the tetragonal basis along the Fe-Fe bond direction. In this paper, we use $x$ and $y$ to denote these Fe-Fe bond directions. The sample was mounted in a custom-built NMR probe incorporating a Razorbill cryogenic strain apparatus \cite{KissikovStrainProbe}.  Uniaxial stress was applied to the crystal as described in \cite{Kissikov2017} by piezoelectric stacks as illustrated in the inset of Fig. \ref{fig:waterfall}, and strain was measured by a capacitive dilatometer. A free-standing NMR coil was placed around the crystal, and spectra were measured
by acquiring echoes while sweeping the magnetic field $H_{0}$ at
fixed frequency. $^{75}$As has spin $I=3/2$, with three separate resonances separated by the quadrupolar interaction. Fig. \ref{fig:waterfall} shows the central and upper transitions as a function of strain at fixed temperature. The higher quadrupolar satellite resonance occurs at field
$H_{sat}=(f_{0}+\nu_{\alpha\alpha})/\gamma(1+K_{\alpha\alpha})$,
where $f_{0} = 55.924$ MHz is the rf frequency, $\gamma=7.29019$ MHz/T is the
gyromagnetic ratio, and $K_{\alpha\alpha}$ and $\nu_{\alpha\alpha}$
are the Knight shift and EFG tensor components in the $\alpha=(x,y,z)$
direction.  The central transition field is given by: $H_{cen}=\frac{f_{0}}{\gamma(1+K_{\alpha\alpha})}\left(\frac{1}{2}+\sqrt{\frac{3f_{0}^{2}-2(\nu_{\beta\beta}+\nu_{\alpha\alpha})^{2}}{12}}\right)$,
where $\beta=(y,x,z)$ for $\alpha={x,y,z}$. The center of gravity
of each peak was used to determine the resonance field, and hence
$K_{\alpha\alpha}$ and $\nu_{\alpha\alpha}$ as a function of strain.  The Knight shift shows essentially no change with strain \cite{Kissikov2017}, however, all components of the EFG tensor show strong variations, as shown in Fig. \ref{fig:nuxx} and Fig. \ref{fig:EFGlinear}.

 The EFG tensor is given by $\nu_{\alpha\beta}=(eQ/12h)\partial^{2}V/\partial x_{\alpha}\partial x_{\beta}$,
where $Q=3.14\times 10^{-29}{\rm m}^2$ is the quadrupolar moment of the $^{75}$As and $V$ is
the electrostatic potential at the As site. This quantity is dominated
by the occupation of the As $4p$ orbitals, which in turn are hybridized with the $d_{xz,yz}$-orbitals of the neighboring
Fe atoms \cite{IyeJPSLorbitalnematicity2015}.
In the tetragonal phase the  EFG asymmetry parameter $\eta=(\nu_{yy}-\nu_{xx})/(\nu_{xx}+\nu_{yy})$ vanishes because the As $4p_x$ and $4p_y$ orbitals are degenerate (i.e. $\nu_{xx} = \nu_{yy}$), as seen in Fig. \ref{fig:nuxx}.  In the presence of nematic order, the $C_4$ symmetry of the EFG tensor is broken and $\nu_{xx}\neq\nu_{yy}$ \cite{DioguardiPdoped2015}.  Because the in-plane anisotropic strain field, $\varepsilon_{ani} = \frac{1}{2}\left(\varepsilon_{xx} - \varepsilon_{yy}\right)$, with $B_{2g}$ symmetry (in the coordinate system of the tetragonal unit cell) couples bilinearly to the nematic order parameter, $\eta$ responds to strain in the same manner that the magnetization of a ferromagnet responds to a uniform magnetic field \cite{TanatarTensileStressPRB,FisherScienceNematic2012,Kuo2015}. Although the applied uniaxial stress also induces strains corresponding to other elastic modes, due to the finite Poisson ratio  the dominant mode is $\varepsilon_{ani}$, which couples to $\eta$. In our configuration we can only apply $\mathbf{H}_0$ perpendicular to the stress axis, which we denote by $x$. We measure both $\nu_{zz} =\nu_{cc}$ along the $\hat{c}$ axis of the crystal, and $\nu_{yy}$ for $\mathbf{H}_0$  in the basal plane. For the latter case, $\nu_{yy}=\nu_{aa}$ for compressive strain ($\varepsilon_{ani}<0$) and $\nu_{yy} = \nu_{bb}$ for tensile strain ($\varepsilon_{ani}>0$), and $\nu_{xx}(\varepsilon_{ani}) = \nu_{yy}(-\varepsilon_{ani})$.
The EFG thus enables us to identify the zero-strain displacement, $x_{0}$, by the condition $|\nu_{xx}|=|\nu_{yy}|=|\nu_{zz}|/2$. Note that  $\eta$ can exceed unity, since $\nu_{xx} + \nu_{yy} + \nu_{zz} =0$.  Furthermore, in the absence of strain a bulk order parameter in a twinned sample would average to zero, whereas the local order measured by NMR reveals all domains simultaneously \cite{takigawa2008}.

As seen in Fig. \ref{fig:nuxx}, the applied strain significantly alters the local EFG. Just above the structural transition the strained EFG values approach those  in the spontaneously ordered phase in the absence of strain.  Furthermore, the maximum strain levels as measured by the dilatometer reach approximately 60\% of the spontaneous values of the orthorhombicity in the ordered phase \cite{NiBa122SingleXtalDiscoveryPRB}.  Nevertheless, $\nu_{yy}$ remains linear over this range as shown in Fig. \ref{fig:EFGlinear}.  The slope of this response is therefore a measure of the static nematic susceptibility, $\chi_{\rm nem}$.  Similar behavior was observed in elastoresistance \cite{FisherScienceNematic2012}, shear modulus \cite{MeingastNematicSusceptibility2016}, and electronic Raman scattering \cite{Ba122RamanPRL2013}. However, the NMR probes the local nematicity in terms of the different orbital occupations reflected in the EFGs, rather than the bulk response, which can be affected by inhomogeneities. We note that, rigorously, $\chi_{\rm nem}$ is the ``bare" nematic susceptibility, i.e. without the contribution arising from the coupling to the lattice. The bare and renormalized susceptibilities are related by a Legendre transformation.

\begin{figure}
	\centering
	\includegraphics[width=\linewidth]{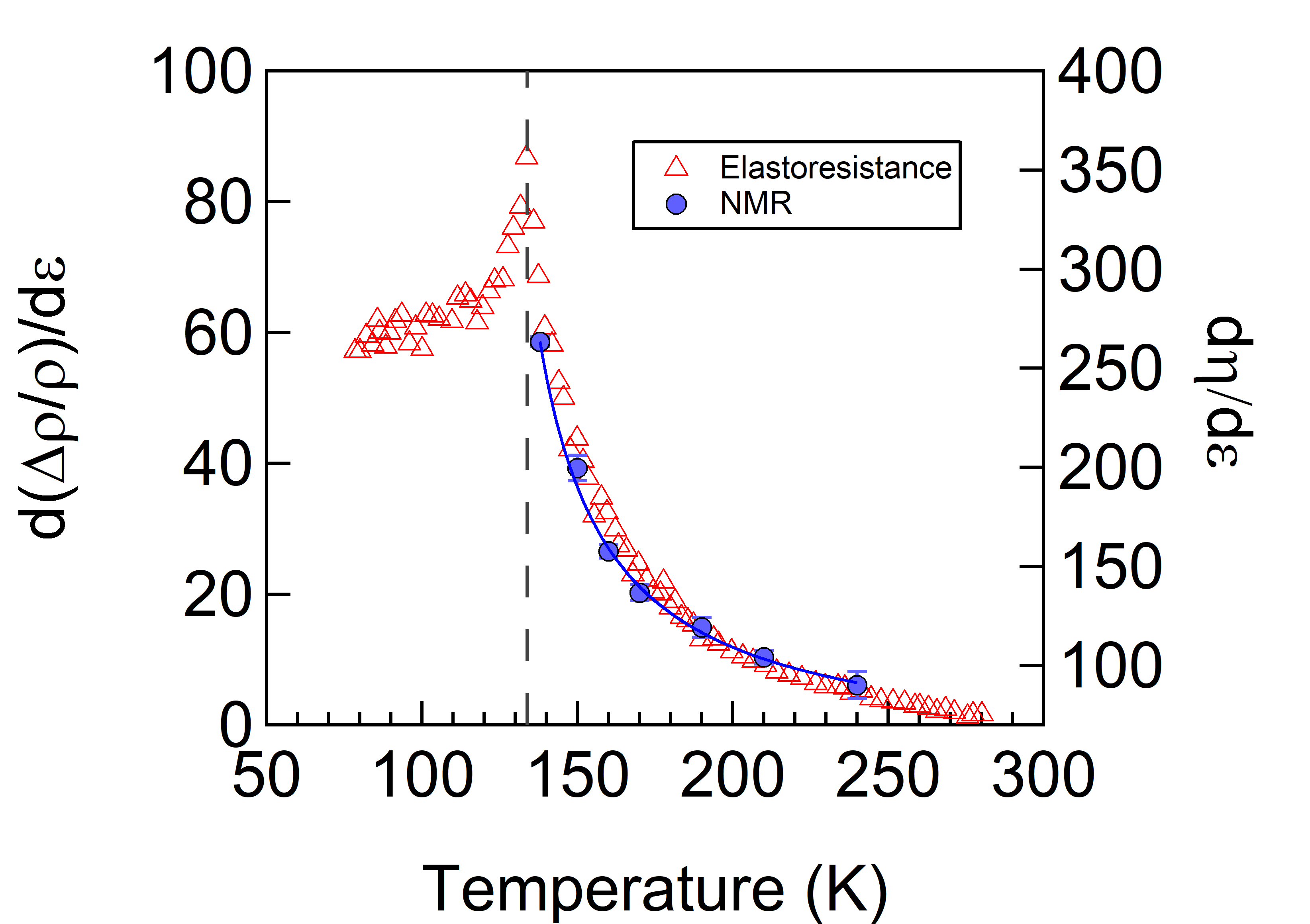}
	\caption{\label{fig:susceptibility} (color online) The nematic susceptibility measured by the EFG asymmetry ($\bullet$) and that measured by elastoresistance ($\vartriangle$, reproduced from \cite{FisherScienceNematic2012}).  The solid line is a fit to the NMR data, as described in the text. The vertical dashed line indicates $T_N$.  }
\end{figure}

Figure \ref{fig:susceptibility} shows the temperature dependence of $d\eta/d\varepsilon_{ani}$ and compares the response to elastoresistance measurements \cite{FisherScienceNematic2012}.   The NMR data exhibit a similar behavior with a divergence at $T_s$.  We fit the EFG data to the sum of a Curie-Weiss term plus a background susceptibility: $\chi_{\rm nem} = C/(T - T_0) + \chi_0$, and find $C_0 = 4700\pm700$ K, $T_0=116\pm3$ K, and $\chi_0 = 54\pm 8$.  The background term reflects the intrinsic response of the lattice, whereas the Curie-Weiss term represents the nematic instability. Our observed value of  $T_0$  is consistent with elastoresistance, but differs from that observed by Raman scattering and by shear modulus measurements \cite{Ba122RamanPRL2013,BlumbergPnictidesNematic2016,MeingastNematicSusceptibility2016}. As noted above, the difference between $T_0$ and $T_s$ arises due to the fact that we are probing the bare nematic susceptibility without the lattice contribution.

In order to understand the relationship between the EFG asymmetry and the splitting between the Fe $d_{xz}$ and $d_{yz}$ orbitals, we have performed {GGA-based DFT calculations} \cite{Blaha2001,Perdew1996} for the tetragonal structure {at 300 K and 0.2 GPa} \cite{Mittal2011} under anisotropic, in-plane strain $\varepsilon_{ani}$. %\textbf{\emph{RMF Why do we have to consider a finite pressure?}}
Our values of the EFG  are consistent with previous calculations in the absence of strain, but underestimate the experimental values by approximately a factor of three \cite{Grafe2009,FurukawaCaKFe4As4}. We confirm that the EFG is dominated by the occupation of the As $p$ orbitals \cite{IyeJPSLorbitalnematicity2015}, which are hybridized with the neighboring $d_{xz}$ and $d_{yz}$ orbitals.  We calculate that $d\eta/d\varepsilon_{ani} = 33$%(21.7 previously)
, which is close to the experimental value of the background susceptibility, $\chi_0$.  The strong temperature-dependent divergence at $T_s$ is a collective phenomenon driven by the electronic system  and cannot be captured by the DFT calculations which are valid only at $T=0$.
Under strain, the energy doublet at the $M$ point in $\mathbf{k}$-space corresponding to the degenerate $d_{yz}$ and $d_{xz}$ onsite energies develop a finite splitting, $\Delta_{xz-yz}$.  
%\textbf{\emph{RMF Is this $n_{xz}-n_{yx}$ or the other way around?}}
We find that $\eta = A\Delta_{xz - yz}$, where $A = 5.7/$eV.  These values are consistent with angle-resolved photoemission experiments that indicate a splitting $\Delta_{xz-yz}\sim 40$ meV in the nematic phase \cite{NaFeAsARPES2012}, whereas NMR studies reveal a value of $\eta \sim 1.2$ \cite{takigawa2008}. 
%\textbf{\emph{RMF Sorry, I'm a bit confused. From Fig. 2 it seems that $\eta$ is approximately $2.5$ in the nematic phase, which would be one order of magnitude larger than the estimate we give}}

Fig. \ref{fig:nuxx} also shows the response of the quadrupolar splitting $\nu_{zz}$ along the c-axis to in-plane strain.   This independent component of the EFG tensor does not couple to the nematic order, but nevertheless it is suppressed by the lattice distortion. We find that $|\nu_{zz}(\varepsilon_{ani})/\nu_{zz}(0)| = 1 - \beta\varepsilon_{ani}^2$, where $\beta \approx 9000$ is approximately temperature independent. Our DFT calculations reveal a small quadratic suppression
%$|\nu_{zz}(\varepsilon_{ani})/\nu_{zz}(0)| = 1 - \beta\varepsilon_{ani}^2$, where
with $\beta = 30$, due to changes in the relative occupations of the As $p_z$  and $p_{x,y}$ orbitals. The difference between the experimental and theoretical values may reflect changes to the c-axis lattice parameters due to a finite Poisson ratio.

%It is likely that response reflects the change of the orbital occupations of the As $4p_z$ orbital under strain.  DFT calculations give $(d\nu_{zz}/d\epsilon)/\nu_{zz}(0) =-1.1$.  Experimentally we find this value is $\approx -40$.  (I wonder how well the DFT calculations get the elastic constants?) Since $\nu_{zz}$ is likely coupled to c-axis lattice parameter, the changes we observe may reflect changes due to a finite Poisson ratio of our sample. (?)

Our measurements offer insight into the behavior of the EFG in electron-doped pnictides.  In doped Ba(Fe,M)$_2$As$_2$ (M = Co, Ni), the quadrupolar satellite resonances are inhomogeneously broadened ($\sim 1.0-1.5$ MHz) relative to those in the parent compound (0.13 MHz) \cite{ImaiLightlyDoped,Dioguardi2010,Takeda:2014ia}.  A large source of this broadening may arise from local strain distributions.  Local strains at dopant atoms can reach up to 3\% \cite{LoucaPRB2011}, which would correspond to a shift in the As EFG parameters of $\delta\eta \sim 10$ and $\delta\nu_{zz} \sim 2.9$ MHz at 140 K. The strain field  relaxes with distance from the dopant, and possibly other types of defects, giving rise to a distribution of local EFGs.   Recently  a finite EFG asymmetry $\eta \sim 0.1$ was reported in BaFe$_2$(As$_{1-x}$P$_x$)$_2$ in the nominally tetragonal phase \cite{IyeJPSLorbitalnematicity2015}. This value would be consistent with an average strain field on the order of $0.05\%$. We postulate, therefore, that the origin of the finite nematicity observed in this compound reflects inhomogeneous strain fields, rather than intrinsic nematicity above the structural transition \cite{Kasahara2012}. The presence of strain fields in the nominally tetragonal phase has indeed been observed directly by STM \cite{Rosenthal2014}. Complex EFG distributions have also been reported in RFeAsO$_{1-x}$F$_x$ (R = La, Sm) that have been interpreted as nanoscale electronic order \cite{GrafeNanoscaleIronPnictides}.  It is unclear whether these spatial variations arise due to $\nu_{zz}$ or $\eta$, although they may reflect a combination of both strain and/or orbital occupations.

In conclusion, we have conducted detailed measurements of the EFG under a uniform uniaxial strain, and observed a linear response that is strongly temperature dependent.  The slope agrees well with other measurements of the nematic susceptibility, and demonstrates that $C_4$ symmetry is broken  not only in the different Fe 3d orbital occupations, but also in the As 4p orbitals.   Our results further demonstrate that $^{75}$As NMR is sensitive to the charge degrees of freedom, and enable a quantitative measure of the local orbital occupations of the Fe d-orbitals. Measurements of the local nematicity by NMR provide an important microscopic complement to other techniques, and offer a unique opportunity to measure the response in the superconducting state. For example, in contrast to elasto-resistance and Raman scattering, NMR under strain can probe the nematic susceptibility below $T_c$.  Such measurements may provide insight into the role of nematic degrees of freedom in the superconducting mechanism \cite{KivelsonNematicQCP2015,Kang2016}.

We thank S. Hillbrand, K. Delong, D. Hemer and P. Klavins, for assistance in the laboratory, and E. Carlson and I. R. Fisher for stimulating discussions. Work at UC Davis was supported by the NSF under Grant No.\ DMR-1506961 (T.K., M.L., B.T.B., N.J.C).  NSF grant DMR-1607139 (W.F.G), DOE NNSA grant DE-NA0002908 grant (W.E.P.).
R.M.F. is supported by the U. S. Department of Energy, Office of Science,
Basic Energy Sciences, under award number DE-SC0012336. R.S. was partially
supported by the DFG through SFB 1143 for the project C02.  Work done
at Ames Lab (S.L.B., P.C.C., M.T., R.P., E.I.T.) was supported by the U.S. Department
of Energy, Office of Basic Energy Science, Division of Materials Sciences
and Engineering. Ames Laboratory is operated for the U.S. Department
of Energy by Iowa State University under Contract No. DE-AC02-07CH11358.

%\bibliography{CurroBibliography}
\bibliography{EFGnematicityBibliography}

\end{document}